# A Continuous Liveness Detection for Voice Authentication on Smart Devices

Linghan Zhang[1], Sheng Tan[1], Yingying Chen[2], and Jie Yang[1]

[1]Florida State University; [2]Rutgers University

**Abstract**—Voice biometrics is drawing increasing attention as it is a promising alternative to legacy passwords for user authentication. Recently, a growing body of work shows that voice biometrics is vulnerable to spoofing through replay attacks, where an adversary tries to spoof voice authentication systems by using a pre-recorded voice sample collected from a genuine user. To this end, we propose VoiceGesture, a liveness detection solution for voice authentication on smart devices such as smartphones and smart speakers. VoiceGesture detects a live user by leveraging both the unique articulatory gesture of the user when speaking a passphrase and the audio hardware advances on these smart devices. Specifically, our system re-uses a pair of built-in speaker and microphone on a smart device as a Doppler radar, which transmits a high-frequency acoustic sound from the speaker and listens to the reflections at the microphone when a user speaks a passphrase. Then we extract Doppler shifts resulted from the user's articulatory gestures for liveness detection. VoiceGesture is practical as it requires neither cumbersome operations nor additional hardware but a speaker and a microphone commonly available on smart devices that support voice input. Our experimental evaluation with 21 participants and different smart devices shows that VoiceGesture achieves over 99% and around 98% detection accuracy for text-dependent and text-independent liveness detection, respectively. Results also show that VoiceGesture is robust to different device placements, low audio sampling frequency, and supports medium range liveness detection on smart speakers in various use scenarios like smart homes and smart vehicles.

## 1 INTRODUCTION

Voice biometrics has been gaining increasing popularity and significance as an alternative to legacy passwords for user authentication on smart devices. Particularly, on mobile devices, voice biometrics has the advantage of natural integration with passwords and face biometrics for multifactor authentication, and thus has been widely adopted by a range of applications for secure access, login, e-commerce and banking [30, 31]. For example, Google develops "Trusted Voice" for Android device access [9]. Whereas, Saypay, a biometric authentication solution provider, supports voice biomet- rics secured online transactions [7]. Furthermore, on the burgeoning smart devices, like smart speakers in homes, workplaces, and smart vehicles, voice biometrics enable secure interactions between the user and various devices, applications, and services. For instance, voice biometrics has been used for IoT device access control to smart speakers, locks, vacuums, fans, and thermostats in the smart home hub [1]. The Enterprise Bank, pairing with ENACOMM's virtual personal assistant, deploys voice biometrics to secure complex personalized banking operations, such as checking balance, making transfers, sending bank statements [2]. Moreover, automakers such as BMW, Audi, Buick, etc. support in-car voice assistants [3] and provide voice-controlled skills like navigation and playing audiobooks and news. Whereas self-driving vehicles makers like Tesla [5] have gone a further step by allowing voice biometrics-based car controls.

However, a growing body of research has demonstrated the vulnerability of voice authentication systems to replay attacks [13], where an adversary tries to spoof the authentication system by using a pre-recorded voice sample collected from a genuine user. The replay attacks are easy to carry out, requiring neither sophisticated equipment nor specific expertise. They are also increasingly practical due to the wide availability of low-cost, high-quality recording and playback devices. The popularity of social media further makes it relatively easy for an adversary to obtain voice samples from the intended target user. Importantly, such low-cost and low-effort attacks are highly effective in spoofing voice authentication systems. For instance, simply replaying a pre-recorded voice command of a user could unlock her/his mobile devices that have the voice-unlock feature (e.g., Android devices) [9]. An extensive study in 2017 shows that replay attacks increase the equal error rate (EER) of state-of-art voice authentication systems from 1.76% to surprisingly 30.71% [13]. Replay attacks thus pose serious threats to the voice authentication systems and have drawn much attention recently.

- L. Zhang and J. Yang are with the Department of Computer Science, Florida State University, Tallahassee, FL, 32306.
  E-mail: {lzhang, jieyang}@cs.fsu.edu
- S. Tan is with the Department of Computer Science, Trinity University, San Antonio, TX, 78212.
  E-mail: stan@trinity.edu
- Y.Chen is with the Department of Electrical and Computer Engineering, Rutgers University, New Brunswick, NJ, 08901.
  E-mail: yingche@scarlet.rutgers.edu

To defend against replay attacks, liveness detection systems distinguish legitimate voice samples of live users from the replayed ones. Traditional methods mainly rely on the acoustic characteristics of an input utterance. Such methods have limited effectiveness in practice as recordings took in benign acoustic environments with high-quality recorders are indistinguishable from the genuine ones [13]. Whereas current voice authentication service providers like Nuance [8] mainly rely on the challenge-response based approaches for liveness detection. Such methods are cumbersome as they require extra user cooperation besides the standard authentication process. Moreover, many researchers target detecting human vital signs associated with speech production for liveness detection. For example, Zhang *et al.* measure the time-difference-of-arrival(TDoA) of different phonemes to the two microphones of the smartphone [28]. Wang *et al.* [24] examine human breathing patterns while Shang *et al.* [20] detect human body vibrations during speech production. These solutions, however, exert strict constraints to users' distances to the devices, i.e., the user needs to hold the device close to or even in touch with their body parts. Besides, a series of liveness detection systems require extra devices for liveness detection. For instance, WiVo [17] and REVOLT [19] use Wi-Fi signals to detect mouth motions and breathing during the human speaks. VAuth [11] and VocalPrint [16] leverage wearable sensors and mmWave to measure the human body vibrations caused by speech production, respectively. Furthermore, many solutions only support text-dependent liveness detection [20], [24], [27], [28] and hence could not protect the conversational voice assistants on the latest smart devices.

In this paper, we introduce VoiceGesture, a continuous liveness detection system that achieves the best of both worlds - i.e., it is highly effective in detecting live users, but does not require the users to perform any cumbersome operations. In particular, our system achieves around 1% EER and above 99% liveness detection accuracy on both average smartphones and smart speakers. Moreover, our solution supports the user to keep their habitual ways of interacting with these devices, i.e., hold the smartphone by ears or in front of the mouth, and speak to a smart speaker from a distance.

Our system measures a user's articulatory gestures when speaking a passphrase for liveness detection. Human speech production relies on the precise, highly coordinated movements of multiple articulators (e.g., the lips, jaw, and tongue) to produce each phoneme sound. It is known as articulatory gesture, which involves multidimensional movements of multiple articulators. Unlike the human, the loudspeaker produces sound relying solely on the diaphragm that moves in one dimension (i.e., forward and backward). Thus, by sensing the articulatory motions when speaking a passphrase, a human speaker can be distinguished from a loudspeaker. Moreover, there exist minute differences in articulatory gestures among people due to individual diversity in the human vocal tract (e.g., shape and size) and the habitual way of pronouncing phoneme sounds [18]. Such minute differences could be further leveraged to detect an adversary who tries to mimic the articulatory gesture of a genuine user.

Therefore, our system re-uses a pair of built-in speaker and microphone on a smart device as a Doppler radar to sense user-specific features of his/her articulatory gestures when the user speaks to the device. In particular, during the enrollment and authentication process, the built-in speaker transmits a high-frequency probe signal at 20kHz [33]. The user's moving articulators reflect this probe signal and cause Doppler frequency shifts at around 20kHz. Meanwhile, the built-in microphone keeps listening and recording the reflections and the user's voice samples. Our system then separates the voice samples for conventional voice authen- tication and extracts individual features from the frequency shifts for liveness detection. In particular, we extract both the frequency and the energy features to reveal the velocity and the location information of the articulatory gestures respectively. To evaluate the performance of our system, we conduct experiments with 21 participants, three differ- ent types of phones, and a smart speaker under various experimental settings. Experimental results show that our system is highly effective in detecting live users and works with users' habitual ways of interacting with the device. The contributions of our work are summarized as follows.

- We leverage smart devices' audio hardware to sense the articulatory gesture of a user when he/she speaks a passphrase. We also show that it is feasible to capture the minute differences in different people's articulatory gestures when they speak the same phoneme sounds.
- We develop VoiceGesture, a liveness detection system that could serve both text-dependent and text-independent voice authentication systems. VoiceGesture extracts user-specific features in the Doppler shifts resulted from the user's articulatory gestures when he/she speaks. VoiceGesture is practical as it requires neither cumbersome operations nor additional hardware other than a pair of speaker and microphone that are commonly available on the latest smart devices.
- Our extensive experimental results show that VoiceGesture achieves over 99% detection accuracy at around 1% EER. Results also show that VoiceGesture can work with different smart devices and sampling frequencies. Especially, with microphone arrays and the beamforming technique, VoiceGesture supports medium-range liveness detection in various use scenarios like smart vehicles and smart homes.

The remainder of the paper expands on the above contributions.

## 2 PRELIMINARIES

### 2.1 System and Attack Model

Voice authentication is the process of verifying a user's claimed identity by extracting the acoustic features that reflect both behavioral and physiological characteristics of a user. This work primarily focuses on the text-dependent systems, as shown in Figure 1, where a user-chosen or system prompted passphrase is used for authentication. We also extend this solution to text-independent liveness detection. Text-dependent systems offer high authentication accuracy with shorter utterances. Therefore, text-dependent authentication is more suitable if security is the highest priority. [25]. However, text-dependent authentication only provides one-time identity verification and only supports enrolled specific passphrases. In comparison, text-independent authentication can enroll and verify the user's identity transparently and continuously regardless of the speech contents. This feature is especially critical to popular conversational voice assistant systems like Google Assistant.

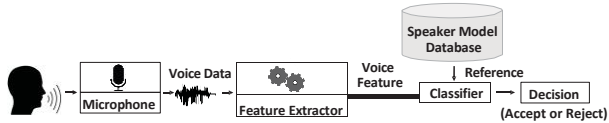

**Fig. 1: A typical text-dependent authentication system.**

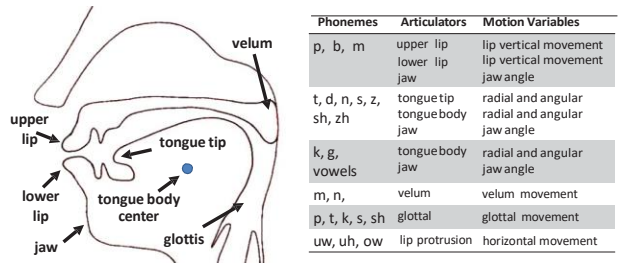

| Phonemes | Articulators | Motion Variables |
|---|---|---|
| p, b, m | upper lip | lip vertical movement |
| | lower lip | lip vertical movement |
| | jaw | jaw angle |
| t, d, n, s, z, | tongue tip | radial and angular |
| sh, zh | tongue body | radial and angular |
| | jaw | jaw angle |
| k, g, | tongue body | radial and angular |
| vowels | jaw | jaw angle |
| m, n, | velum | velum movement |
| p, t, k, s, sh | glottal | glottal movement |
| uw, uh, ow | lip protrusion | horizontal movement |

**Fig. 2: Articulators, phonemes and the corresponding articulatory gestures.**

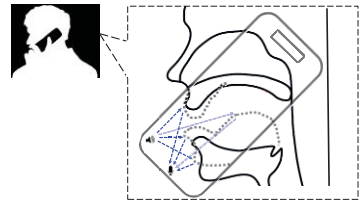

**Fig. 3: An illustration of sensing the articulatory gesture when a user speaks a passphrase on the phone.**

For the attack models, we consider replay attacks in our work as they are easy to implement and highly effective in spoofing the voice authentication systems. In particular, we consider two types of replay attacks: *playback attack* and *mimicry attack*. In a playback attack, an adversary uses a loudspeaker to replay a pre-recorded passphrase of an intended target user. Given that attackers may know the defending strategy of our liveness detection system, they could conduct more sophisticated mimicry attacks, in which an adversary tries to mimic the articulatory gestures of a genuine user. To perform a mimicry attack, the adversary can use a far-field speaker to replay a pre-recorded passphrase and simultaneously mimic the victim's articulatory gesture corresponding to the replaying passphrases. In mimicry attacks, we also consider that the attacker can observe how a genuine user pronounces the passphrase, for example, by taking a video of the genuine user, and then practice before conducting the attack.

### 2.2 Articulatory Gesture

Human speech production requires precise and highly coordinated movements of multiple articulators. Specifically, articulatory gestures are used to describe the connection between the lexical units with the articulator dynamic when producing speech sounds. For English speech production, the coordination among multiple articulators produces gestures like lip protrusion, lip closure, tongue tip, tongue body constriction, and jaw angle. For example, three articulators including upper lip, lower lip, and jaw are involved when a speaker conducts the gesture of lip closure, which could lead to the phoneme sounds of *[p]*, *[b]* and *[m]*.

Figure 2 illustrates various articulators and their locations, common English phonemes, and the corresponding articulatory gestures. Each phoneme sound production usually involves multidimensional movements of multiple articulators. For instance, the pronunciation of the phoneme *[p]* requires upper and lower lips horizontal movements and jaw angle change. Moreover, although some phonemes share the same articulatory gestures, their movement speeds and intensities could be different. For example, both *[d]* and *[z]* require tongue tip constrictions. However, they differ in terms of the exact tongue tip radial and angular position.

### 2.3 Sensing the Articulatory Gesture

We leverage the Doppler effect to sense the articulatory gestures. Figure 3 shows one example of sensing the articulatory gestures when a user speaks a passphrase to a phone held by his/her ear. The built-in speaker of the phone emits a high-frequency tone, which is reflected by multiple articulators of the user. The reflections are then recorded by the built-in microphone of the same phone. In our context, the articulators reflecting the probe signal can be considered as virtual transmitters that generate the reflected sound waves. As the articulators move towards the microphone, the crests and troughs of the reflected sound waves arrive at the microphone at a faster rate. Conversely, if the articulators move away from the microphone, the crests and troughs arrive slower. In particular, an articulator moving at a speed of $v$ with an angle of $\alpha$ from the microphone results in a Doppler shift (i.e., frequency change $\Delta f$) of:

$$\Delta f \propto \frac{v \cos(\alpha)}{c} f_0, \qquad (1)$$

where $f_0$ is the frequency of the transmitted sound wave and $c$ is the speed of sound in the medium.

We observe from Equation (1) that a higher frequency of the emitted sound (i.e., $f_0$) results in a larger Doppler shift for the same articulator movements. We thus choose to emit a high-frequency sound at 20kHz, which is close to the limit of the built-in speaker/microphone of the phone. Such a high-frequency signal maximizes the Doppler shifts caused by the articulatory gesture and is also inaudible to the human ear. Moreover, the Doppler shifts are vectors decided by the moving directions of the articulators (i.e., $\alpha$). An articulator moving away from the microphone results in negative Doppler shift, while an articulator moving towards the microphone leads to a positive Doppler shift. As each phoneme pronunciation involves multidimensional movements of multiple articulators, the resulted Doppler shifts at the microphone are a superposition of sinusoids at different shifts. In addition, a faster speed (i.e., $v$) results in a larger Doppler shift. The magnitude of the Doppler shift thus can be further utilized to distinguish different gestures or people that produce the same phoneme sound with various speeds. Furthermore, the reflections from the articulators that closer to the microphone result in stronger energy due to the signal attenuation in the medium. Therefore, the Doppler shifts' energy distribution provides another dimension of information for differentiating articulatory gestures.

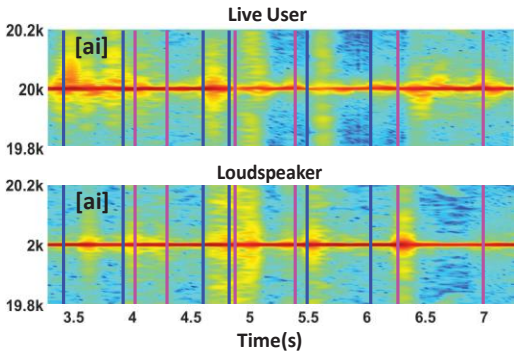

**Fig. 4: Doppler shifts of a live user and a speaker replay.**

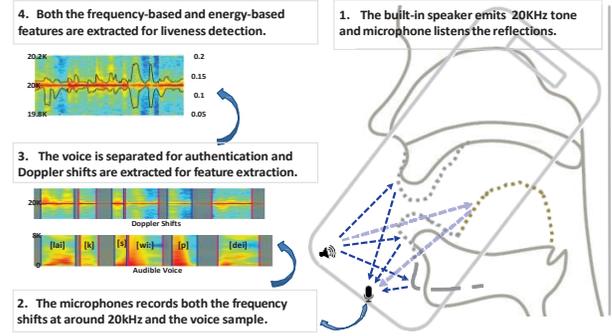

**Fig. 5: Illustration of the articulatory gesture based liveness detection on smartphone.**

### 2.4 Loudspeaker

Unlike the human, loudspeaker relies on solely the diaphragm that moves in one dimension to produce sound wave. Specifically, the loudspeaker diaphragms move forward and backward to increase and decrease the air pressure in front of it, thus creating sound waves. Such movements are controlled by the frequency and intensity of the input audio signal. For instance, the input sound that possesses a high pitch results in fast movement of the diaphragm, while when a user turning up the volume, the diaphragm pushes harder to produce a higher pressure in the air.

A loudspeaker could be distinguished from a live speaker based on the movement of articulators. First, they differ in terms of the movement complexity and the number of the articulators. Besides, the movements of the human articulators do not always produce sound, whereas the movements of diaphragms certainly result in sound waves. Figure 4 shows the Doppler shifts sensed by the probe signal at 20kHz when a loudspeaker replays and a live user speaks, respectively. The frequency distribution inside each pair of vertical bars in the figure corresponds to the Doppler shifts resulted from one same phoneme sound. We could observe that the Doppler shifts of the loudspeaker look relatively clean due to the simple diaphragm movements, whereas the Doppler shifts caused by a live user's complex articulatory gestures spread out over a much larger volume of space.

### 2.5 Individual Diversity of Articulator Gesture

There exist minute differences in articulatory gestures among people when producing the same phoneme due to the individual diversity in the human vocal tract and the habitual way of pronunciation. For example, research shows that different people adopt different movement trajectories of articulators to produce the same utterance [15]. The physiological features of the vocal tract, such as the sizes and shapes of the lips and tongues, vary among people [21]. Moreover, there are diverse articulatory strategies for sound production. For instance, some speakers' jaw movements are closely connected with tongue body gestures, while others are not [12].

According to research on five individuals [14] articulatory gestures, the averaged speed differences of their upper lips and jaws are 0.04m/s and 0.06m/s, respectively. Given the duration of a phoneme sound is around 250ms and most smart devices support 192kHz sampling frequency, we could achieve 1Hz frequency resolution when calculating each phoneme's frequency shifts. Moreover, with the 20kHz probe signal, 1Hz Doppler shift corresponds to the articulator speed of 0.017m/s, which provides a much higher sensitivity than that of the speed difference in both upper lip and jaw movements (i.e., 0.04m/s and 0.06m/s). We thus enable to differentiate different people even if they are pronouncing the same phoneme sound. The differences in articulatory gestures are expected to be much smaller under the mimicry attacks, where an adversary mimics the articulatory gesture of a genuine user. Nevertheless, each articulatory gesture involves movements of multiple articulators that provide more information to detect the attacks. In addition, a passphrase consists of a sequence of phoneme sounds, which dramatically increase the possibility to distinguish between a genuine user and an attacker.

## 3 SYSTEM DESIGN

### 3.1 Approach Overview

The key idea underlying our liveness detection system is to leverage the smart devices' audio hardware to sense the articulatory gesture of a sequence of phoneme sounds when a user speaks to the devices. Taking the smartphone use case in Figure 5 as an example, the built-in speaker at the bottom of the phone starts to emit an inaudible acoustic tone at 20kHz once the authentication system is triggered. When a user speaks a passphrase, the built-in microphone records user's voice as well as the inaudible acoustic tone and its reflections. Then we extract features based on both the frequency shift and energy distribution in the observed Doppler shifts around 20kHz, and compared those against the passphrase-based or phoneme-based user profile obtained during user enrollment for text-dependent or text-independent liveness detection. A live user is declared if the similarity score exceeds a predefined threshold. Under playback attacks, the extracted Doppler shift features are different from the user profile due to the fundamental difference between the human speech and the loudspeaker sound production systems. Under mimicry attacks, the extracted features show minute differences from the user profile given individual diversity of human vocal tract and the habitual ways of pronunciation.

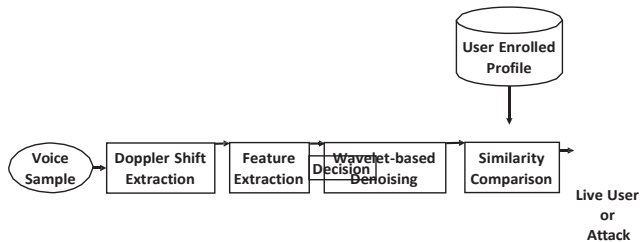

**Fig. 6: The flow of our liveness detection system.**

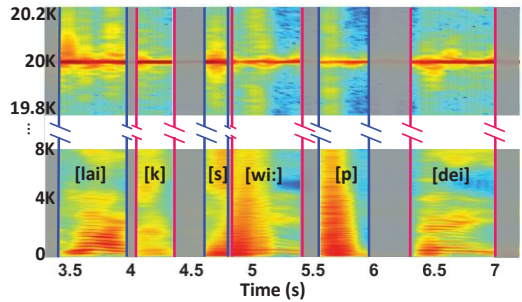

**Fig. 7: An illustration of Doppler shifts extraction based on phoneme sounds.**

Our system works when the users hold the phones with their nature habits as opposed to the prior smartphone based solutions that require users to hold or move the phone in some predefined manners. Moreover, our system support medium-range (up to 1m) text-independent liveness detection for smart device use cases in IoT environments like smart vehicles and smart homes. Comparing with the commercially used challenge-response based solutions, our system does not require any cumbersome operations besides the standard authentication process. Once it integrated with a voice authentication system, the liveness detection is totally transparent to the users.

### 3.2 System Flow

Realizing our system requires five major components: *Doppler Shifts Extraction*, *Feature Extraction*, *Wavelet-based Denoising*, *Similarity Comparison*, and *Detection*. As shown in Figure 6, the acoustic signal captured by the built-in microphone first passes through the Doppler Shifts Extraction process, which extracts the Doppler shifts for each phoneme sound in the spoken utterance. We rely on the user's audible voice samples to separate each phoneme and the corresponding Doppler shifts. Then, we map the segmentation to the inaudible frequency range at around 20kHz frequency to extract the Doppler shifts of each phoneme. Next, the Feature Extraction component extracts both energy-band and frequency-band features from the Doppler shifts. Then we utilize wavelet-based denoising technique to further remove the mixed noises. At last, our system matches the frequency-based and energy-based features with the ones stored in the liveness detection system by using cross-correlation coefficient. It yields a similarity score, which is compared against a predefined threshold. If the score is higher than the threshold, a live user is detected. Otherwise, an attack is declared.

### 3.3 Doppler Shifts Extraction

Once finish recording, our system first separates the user's voice samples (i.e., below 10kHz) for conventional voice authentication. Then, we rely on the audible voice sample to separate each phoneme and the corresponding Doppler shifts at around 20kHz. Specifically, we convert the recorded signal from the time domain to the frequency domain by performing Short-Time Fourier Transform (STFT) with a window size of 250ms. Figure 7 shows one example of the spectrogram of the recorded signal when a user speaks "*like sweep day*". We can find that the audio voice sample is less than 10kHz, and the Doppler shifts are usually within 200Hz at around 20kHz. Such a large gap ensures the voice sample will not be affected by the high frequency of 20kHz and its Doppler shifts. Given the spectrogram of the recorded signal, we aim to extract the Doppler shifts for each individual phoneme. For text-dependent liveness detection, we further remove the pauses due to transaction between phoneme sounds and also the transactions between words (i.e., the shaded bars in the figure).

Then we perform phoneme segmentation [28] to obtain segmented and labeled phonemes for each word. Finally, our system matches the time stamp of each phoneme segmentation to 20kHz frequency range to extract the corresponding Doppler shifts. One example is shown in Figure 7, which illustrates six segmented phoneme sounds (i.e., [*lai*], [*k*], [*s*], [*wiː*], [*p*], [*dei*]) and the corresponding Doppler shifts at around 20kHz. We observe that the phonemes like [*lai*] and [*dei*] display more intensive Doppler shifts than these of the phonemes like [*k*] and [*p*]. This is because when pronouncing [*lai*], larger movements from multiple articulators, including lips, jaw, and tongues, are required. In contrast, when pronouncing [*p*], only small movements from lips and jaw are involved.

### 3.4 Feature Extraction

After we obtain the Doppler shifts of all the phonemes, we first normalize them as the same length as those in the user profile. Such normalization is used to mitigate the user's different speech speeds when performing voice authentication. Then, we re-splice the normalized Doppler shifts of each phoneme together for text-dependent liveness detection. To eliminate the interferences due to other movements such as nearby moving objects or body movements, we further utilize a Butterworth filter with cut-off frequencies of 19.8kHz and 20.2kHz to remove these out-of-band noises.

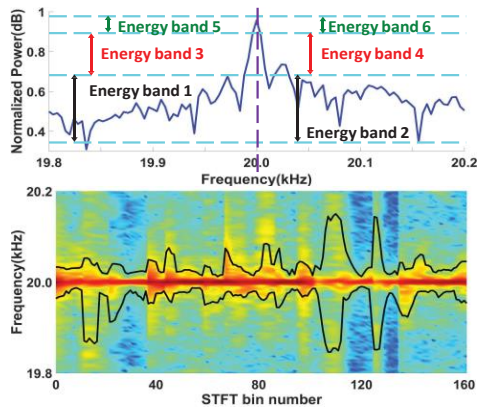

**Fig. 8: An example of energy sub-band and energy-based frequency contours.**

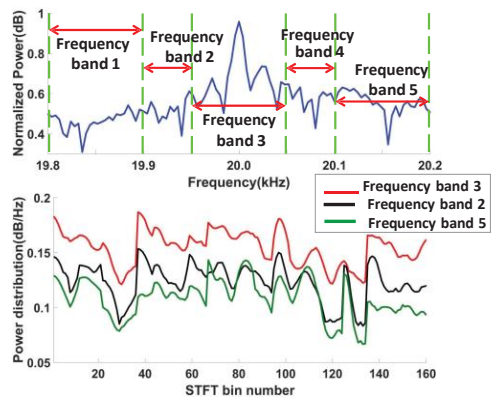

**Fig. 9: An example of frequency sub-band and frequency-based energy contours.**

Next, we extract two types of features from the Doppler shifts of the whole passphrase (text-dependent systems) or single phonemes (text-independent systems): energy-band frequency features and frequency-band energy features.

The first type of feature quantifies the relative movement speeds among multiple articulations. By dividing the energy level of all the frequency shifts into several different bands, we can separate different parts of articulators based on their distances to the microphone. With higher energy of the captured Doppler shifts, a closer movement occurred with respect to the microphone. Before energy band partition, we first normalize each segmented phoneme's energy level into the same scale (i.e., from 0 to 1). Such normalization can mitigate the energy shifts caused by the inconsistency of a user speaking the same utterance to the device. We partition the energy into three levels based on the energy distribution, resulting in 6 sub-bands. Each energy level includes both positive and negative Doppler shifts, as shown in the top graph of Figure 8. Sub-band 5 and 6 with power levels between 0.95 to 0.99 capture the strongest Doppler shifts that resulted from closest articulators like lips. Sub-band 3 and 4 include the power level from 0.7 to 0.9. They catch the Doppler shifts caused by closer articulators like jaws. Whereas sub-band 1 and 2 with lowest energy level between 0.4 to 0.7 contain Doppler shifts of the farthest articulators like the tongue. Given each sub-band, we use the centroid frequency as the feature and combine all the centroid frequencies of each phoneme, resulting in one frequency contour for each band. The bottom part of Figure 8 demonstrates two energy-band frequency contours (i.e., band 1 and 2) extracted from the sentence "*Oscar didn't like sweep day*" spoken by a live user. Those two bands represent articulators (e.g., the tongue) with a longer distance to the microphone. In this Figure, we could observe that the frequency shifts vary considerably when the user pronounces different phonemes.

The second type of feature is the frequency-band energy feature, which quantifies the relative movement positions among multiple articulations across phonemes. As a faster movement results in a larger magnitude of Doppler shift, we can compare the energy levels of different articulators moving at similar velocities. In particular, we divide the frequency shifts into 5 major sub-bands considering three levels of velocities in both positive and negative directions, as shown in the upper part of Figure 9. Sub-band 3 covers frequency shifts from -50Hz to 50Hz, sub-band 2 and 4 include frequency shifts from 50Hz to 100Hz and -100Hz to -50Hz, and sub-band 1 and 5 correspond to 100Hz to 200Hz and -200Hz to -100Hz. Similar to the frequency contour, we calculate the average energy level at each frequency sub-band, and then splice the resulted energy levels together to form an energy contour. The lower part of Figure 9 demonstrates three frequency-band energy contours at the band 2, 3, and 5. We observe that the frequency band 3 contour has the highest energy level. It is because while speaking an utterance, the lower facial region of a user moves slowly. Nevertheless, the large size of the lower facial region leads to strong signal reflections. The frequency band 5 contour demonstrates the lowest energy level caused by articulators farthest from the microphone, such as the tongue.

### 3.5 Wavelet-based Denoising

The purpose of wavelet-based denoising is to remove the noisy component mixed in the extracted features. Those components could be caused by hardware imperfections or surrounding environment interferences and noises. Thus our system utilizes wavelet denoising technique based on Discrete Wavelet Transform (DWT) to further analyze the signal in both time and frequency domain. It decomposes the input signal into two components: approximation coefficients and detailed coefficients. The approximation coefficients depict the trend of the input signal, representing large-scale features. Meanwhile, the detailed coefficients retain the small scale characteristics, which mix both fine details of the signal and noisy components. Our goal is to extract the fine details while removing the mixed noises.

In particular, our system first decomposes each extracted contour into approximation and detailed coefficients by going through low pass and high pass filters. We run this step recursively for 3 levels. After obtaining multiple levels of detailed coefficients, a dynamic threshold is applied to each level of detail coefficients to filter out the mixed noises (i.e., the readings with small values). Then, we combine the original approximation coefficients with the filtered detail coefficients. After that, we use the inverse DWT to reconstruct the denoised contour.

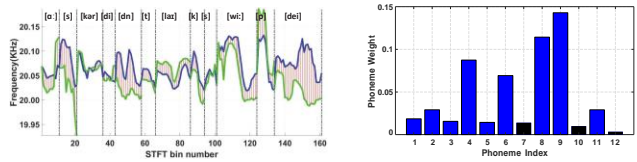

(a) An example of phoneme-based templates for a sentence.

(b) The weights of phoneme-based templates.

**Fig. 10: Weighted Phoneme-based Templates Building.**

## 3.6 Template Building

For text-dependent liveness detection, we build and compare passphrase-based contour features. While for text-independent liveness detection, we establish a set of phoneme-based templates for each individual user. We notice that the contour features of some phonemes are more stable than those of others. For example, a short-sound phoneme tends to provide more consistent contour features than a long-sound phoneme. The reason is that when pronouncing a short-sound phoneme like a consonant or a monophthong, the individual's articulator movements are monotonous. However, when a human pronouncing a long-sound phoneme, especially a diphthong, the articulator movements could be changeable. This observation may vary depending on the speakers' accent.

Therefore, we take advantage of such a phenomenon to improve the system stability by assigning different weights to phoneme templates according to their consistency. This could enhance the impact of the phonemes with stable features, whereas lower the influence of the phonemes with unstable features. Specifically, we align the beginning of each phoneme contour and then adopt the following equation to calculate the weight of a phoneme-based template:

$$w = \frac{\sum_{i=1}^{n} L_i}{\sum_{i,j=1}^{n} (A_i - A_j))} \quad (2)$$

where $A$ is any one of the same type of contour features, and $L$ is the length of the phoneme. The denominator of this equation calculates the aggregate areas between the contours of any two trials of the same phoneme. We remove the impact of the phoneme length by introducing the numerator that computes the total lengths of $n$ trials of pronunciations. Figure 10(b) displays the weights of 12 phonemes in Figure 10(a). As we could note that consonants or monophthongs like [s] [di] yield more consistent contour than diphthongs like [dei].

## 3.7 Similarity Comparison

**Text-dependent Similarity Comparison.** For the text-dependent liveness detection system, to compare the similarity of each extracted contour feature with the corresponding one in the user profile, we use the correlation coefficient technique, which measures the degree of linear relationship between two input sequences. The resulted correlation coefficient ranges from −1 to +1, where the value closer to +1 indicates a higher level of similarity and a value closer to 0 implies a lack of similarity [32].

In particular, given a series of $n$ values in each energy-band frequency or frequency-band energy contour $A$ and the corresponding pre-built user profile $B$, written as $A_i$ and $B_i$, where $i = 1, 2, ..., n$. The Pearson correlation coefficient can be calcualted as:

$$r_{AB} = \frac{\sum_{i=1}^{n}(A_i - \bar{A})(B_i - \bar{B})}{(n-1)\delta_A\delta_B}, \quad (3)$$

where $\bar{A}$ and $\bar{B}$ are the sample means of $A$ and $B$, $\delta_A$ and $\delta_B$ are the sample standard deviations of $A$ and $B$.

**Text-independent Similarity Comparison.** For the text-independent liveness detection, while the user speaks, our system keeps listening while searching for the phoneme-based templates for each phoneme in the speech. After collecting all the templates for the current sentence, we compare the similarity between the contour features of each phoneme in this sentence and the phoneme-based weighted templates with the following equation:

$$\rho_{xy} = \frac{\sum_{i,j=1}^{n}[w_i(A_i - \bar{A}_i)(B_i - \bar{B}_i)]]}{\sqrt{\sum_{i=1}^{n}(w^i(A^i - \bar{A}))^2}\sqrt{\sum_{i=1}^{n}(w^i(B^i - \bar{B}))^2}} \quad (4)$$

where $A_i$ and $B_i$ are the corresponding contours of the $i^{th}$ phoneme, and $w_i$ is the weight of this phoneme. To be noticed, both $A_i$ and $B_i$ are sequences, and $barA_i$ and $barB_i$ are the averages of $A_i$ and $B_i$ respectively. Before comparison, we normalize $A_i$ and $B_i$ to become the same length, then we apply $w_i$ to each point in $A_i$ and $B_i$.

These procedures enable text-independent liveness detection by comparing the current speech with the weighted, phoneme-based templates, rather than the passphrase-based templates. Therefore, our liveness detection system could protect the whole communication session continuously in the IoT environments.

To detect a live user, we use energy-based frequency contours (i.e., energy-based feature), frequency-band energy contours (i.e., frequency-based feature), and combined feature of these two (combined feature), respectively. Given the correlation coefficients of all contours, we simply compare the averaged coefficient to a predefined threshold for live user detection.

## 4 PERFORMANCE EVALUATION

In this section, we present the the experimental performance of our liveness detection system under both *replay* and *mimic* attacks. The project has obtained IRB approval.

### 4.1 Experiment Methodology

**Phones and Placements.** We employ three types of phones including Galaxy S5, Galaxy Note3, and Galaxy Note5 for our evaluation. These phones differ in terms of sizes and audio chipsets. Specifically, the lengths of S5, Note3 and Note5 are 14.1cm, 15.1cm and 15.5cm respectively, whereas the chipsets are Wolfson WM1840, 800 MSM8974 and Audience's ADNC ES704, respectively. All the audio chips and the speaker/microphones of these phones can record and playback 20kHz frequency sound. The operating systems of those phones are the Android 6.0 Marshmallow that released in 2015, which supports audio recording and play back at 192kHz sampling frequency. We thus evaluate our system with the sampling frequencies including 48kHz, 96kHz and 192kHz. We present the results for 192kHz sampling frequency in the evaluation unless otherwise stated. Additionally, we consider two types of phone placements that people usually used to talk on the phone: have the phone held either to user's ear or in front of the mouth.

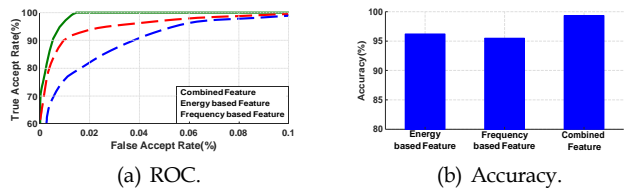

(a) ROC.

(b) Accuracy.

**Fig. 11: Overall Performance**

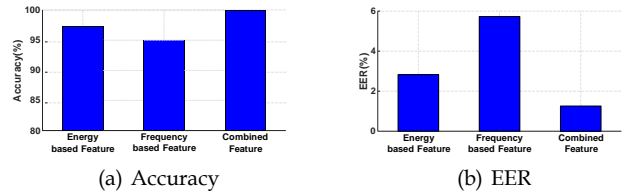

(a) Accuracy

(b) EER

**Fig. 12: Performance under Replay Attacks**

**Smart Speaker Setup** We adopt MiniDSP UMA-8 [6] for experiments on smart speakers as it deploys circular arranged microphone arrays like many state-of-the-art popular smart speakers (e.g., Google Home and Amazon Echo). Moreover, it grants users access to the raw recording data. In particular, MiniDSP UMA-8 has a microphone array composed of 7 MEMS microphones. One of the microphones is located at the center and the other 6 microphones are uniformly configured around a circular board with a radius of 0.43 m. To extend the effective range of our liveness detection solution, we utilize the Delay-and-Sum beamforming technique on the microphone array. The Delay-and-Sum beamforming is based on the fact that the signals received by these microphones are similar, nevertheless, they have different delays and phases. Therefore, it calculates the time difference of arrival of the signals received by the 6 microphones and that recorded by the microphone in the center, and then shifts the signals by corresponding phases and sums them up.

**Data Collection.** Our experiments involve 21 participants, including 11 males and 10 females. The participants are recruited by emails including both graduate students and undergraduate students. These participants include both native and non-native English speakers with ages from 21 to 35. We explicitly tell the participants that the purpose of the experiments is to perform voice authentication and liveness detection. Each participant chooses his/her own 10 different passphrases. For text-dependent liveness detection, they repeat each passphrase three times to enroll in the authentication system and use the averaged features to establish the user profiles. Whereas for text-independent liveness detection, the participants read the "Rainbow Passage" [4] that contains all English phonemes to create the phoneme-based frequency shift templates. Each participant tries 10 times for each passphrase to perform legitimate authentication, which totals 2100 positive cases. The lengths of those passphrases range from 2 to 10 words with one third are 2 to 4 words, one third are 5 to 7 words, and the rest are 8 to 10 words. In addition, to evaluate the individual diversity among users, we ask 12 out of the 21 participants to pronounce the same passphrase. Our experiments are conducted in classrooms, apartments, and offices with background and ambient noises such as HVAC noises and people chatting.

**Attacks.** We evaluate our system under two types of replay attack: *playback attacks* and *mimicry attacks*. Both forms of attacks are considered in our evaluation sections unless claimed otherwise. The playback attacks are conducted with loudspeakers including the standalone speakers, the built-in speakers of mobile devices, and the earbuds. In particular, a DELL AC411 loudspeaker, the built-in speaker of Note5 and a pair of Samsung earbud are used to playback the participants' voice samples in front of the smartphone that performing voice authentication. Specifically, each form of these speakers replays voice samples from 10 participants, and the build-in speaker/earbud and the loudspeaker contributes 3 and 4 trials for each of the 10 passphrases respectively, amounting to 1000 replay attacks. All replay attacks are captured by an identical phone with the same holding position that the participants used for authentication.

For mimicry attacks, we first record the articulatory gesture of the participants when they speaking the passphrase by using a digital video recorder. The video recording only covers the lower facial region for privacy concerns. Such a lower facial region including the articulator movement of upper and lower lips, tongue and jaw. Then other participants are invited to watch the video carefully and repeatedly practice the pronunciation by mimicking the articulatory gesture in the video. In particular, they are instructed to mimic the speed of talking, the intensity and range of each articulator movement, the speech tempo and etc. After they claim that they have learned how the person in the video speaks and moves the articulators, they start to conduct the mimicry attacks in front of the smartphone that used for voice authentication. We recruit 4 attackers and each mimics 6 participants. For each victim/participant, 5 trials for each of 5 passphrases are mimicked. There are in total 600 mimicry attack attempts.

### 4.2 Overall Performance

We first present the overall performance of our system in detecting live users under both playback and mimicry attacks. Figure 11(a) depicts the ROC curves of our system under both types of attacks. We observe that with 1% FAR, the detection rate is as high as 98% when using the combined features. Such an observation suggests that our system is highly effective in detecting live users under both replay and mimic attacks. Moreover, we find that the energy-based feature results in better performance than that of the frequency-based feature. For example, with 1% FAR, the frequency-based feature provide the detection rate at around 90%. Furthermore, we observe that the participants who have smaller scale of articulatory movements generate higher false accept rate. Additionally, Figure 11(b) shows the overall accuracy under both attacks. Similarity, we observe that combined feature has the best performance, with an accuracy at about 99.34%, whereas the energy-based feature alone achieves an accuracy of 96.22%. The time to perform an authentication is about 0.5 seconds on a laptop server. The above results demonstrate the effectiveness of our system in detecting live users. Also, the energy-based feature and frequency-based feature can complement each other to improve the detection performance.

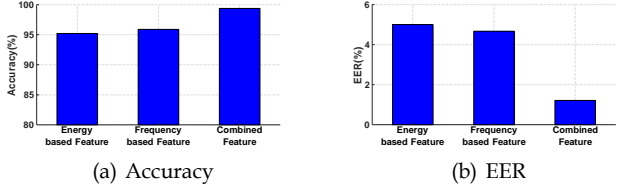

**Fig. 13: Performance under Mimicry Attacks**

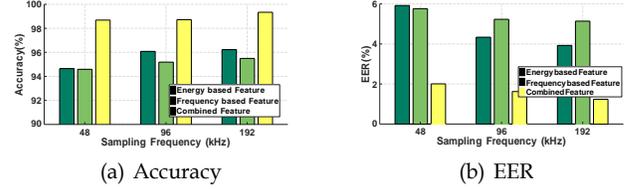

**Fig. 15: Performance under different sampling frequencies**

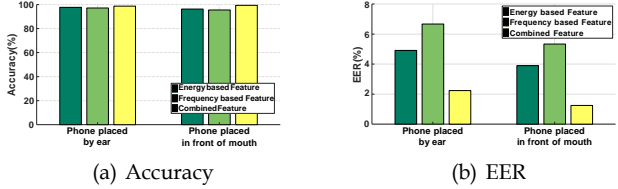

**Fig. 14: Performance under different phone placements**

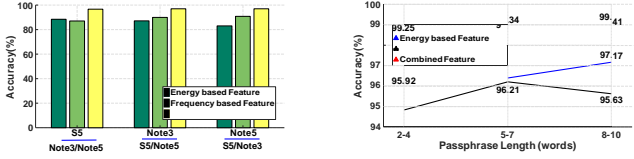

**Fig. 16: Accuracy of using a phone for enrollment and another for authentication.**

**Fig. 17: Accuracy under different length of passphrase.**

**Playback Attack.** We next detail the performance under playback attacks. Figure 12 shows the performance in terms of accuracy and EER under replay attacks. We observe that the combined feature results in the best performance. It has an accuracy of 99.3% and an EER of 1.26%. In particular, with only one type of feature, we can achieve an accuracy of 97.41% and an EER of 2.83%. These results show that the two types of feature can complement with each other and the combined feature is very effective in detecting live user under playback attacks.

**Mimicry Attack.** Next, we study the detailed performance under mimicry attacks. Figure 13 shows both the the accuracy and EER of our system. Again, the combined feature achieves the best accuracy at about 99.3% and an EER of 1.21%. Unlike the playback attack scenario, the frequency-based feature has better performance than that of the energy-based feature. In particular, the frequency-based feature has an accuracy of 95.9% and an EER of 4.67%. The above results suggest that the extracted features from the Doppler shifts of a sequence of phoneme sounds could capture the differences of the articulatory gesture between an attacker and a live user under mimicry attacks. Thus, our system is effective in detecting live users under mimicry attacks.

### 4.3 Impact of Phone's Placement

Different users may have different habits to talk on the phone in terms of how to hold the phone while speaking. We thus compare the performance under two placements of the phone (i.e., hold the phone to ear and hold the phone in front of the mouth) that people usually feel comfortable to use. Figure 14(a) presents the performance comparison of the accuracy, whereas Figure 14(b) shows the comparison of the EER. In high level, the results show that our system is highly effective under both placements. In particular, when placing the phone to the ear, we have the best accuracy as 98.61%, while the best accuracy for placing the phone in front of the mouth is slightly higher. This is due to the fact that placing the phone in front of the mouth can capture the movement of the tongue better as the microphone is directly facing the mouth. Similarly, placing the phone to the ear has slightly worse EER, i.e., at 2.24%, whereas it is about 1.2% for the other placement. Nevertheless, our system works well under both placements and could accommodate different users who have different habits to hold the phone while talking. This property of our system indicates our system doesn't require the user to hold the phone at a specific position or move the phone in a predefined manner as opposed to the prior smartphone based solutions.

### 4.4 Impact of Sampling Frequency

We next show that how well our system can work with some low-end phones that can only playback and record at 48kHz or 96kHz sampling frequency. Figure 15(a) depicts the accuracy of our system under 48kHz, 96kHz and 192kHz sampling frequencies. We notice that a higher sampling frequency results in a better performance. This is because a higher sampling frequency could capture more details of the articulatory gestures and has a better frequency resolution. In particular, the combined feature achieves an accuracy of 98.72% for 96kHz sampling frequency, and 98.69% for 48kHz sampling frequency. Moreover, Figure 15(b) shows the EER under those three sampling frequencies. We find the 96kHz sampling frequency has an EER of 1.63%, whereas it is 2.01% for 48kHz sampling frequency. These results indicate that our system still works very well at a lower sampling frequency. Thus, our system is compatible to those older version smartphones.

### 4.5 Impact of Different Phones

Our system also supports the users to use different types of phones for enrollment and online authentication. Specifically, we experiment with three different phones including

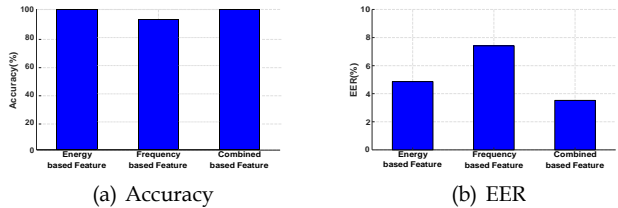

(a) Accuracy      (b) EER

**Fig. 18: Text-independent Liveness Detection Performance.**

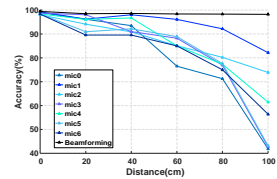

**Fig. 19: Accuracy with smart speaker at different distances.**

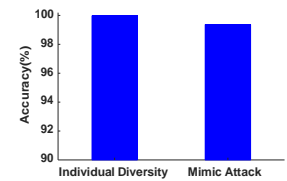

**Fig. 20: Individual diversity v.s. Mimicry attacks.**

S5, Note3 and Note5. In the experiments, the participants use one of these three phones to enroll in the system but use the other two phones for online voice authentication. The performance of our system is in Figure 16. Results show that our system works well under such scenarios. In particular, the combined feature provides an accuracy of 96.58%, 96.93% and 96.98% when using S5, Note3, and Note5 as the enrollment phone, respectively. Results also indicate that the performance is comparably well no mater which phone is used for enrollment. Although the accuracy is slightly worse than that of using the same phone for enrollment and authentication, our system is still able to accommodate different types of phones.

### 4.6 Impact of Passphrase Length

Next, we show how the length of each passphrase affects the performance of our system. Security professionals usually suggest to choose a passphrase with more than 5 words so as to provide a desired security. In the light of this, we classify the passphrases into three categories according to their lengths: 2 to 4 words, 5 to 7 words, and 8 to 10 words. Figure 17 displays the accuracy of our system with different lengths of passphrases. We could observe that when increasing the length of the passphrase, the accuracy slightly improved from 99.25% to 99.41%. This is expected as a longer passphrase results in more articulatory gestures for differentiating a live user from an attacker. Moreover, we observe the improvement is not obvious, since we extract 11-dimensional features from each phoneme, which suggests that 2 to 4 words passphrases containing around 10 to 20 phonemes could provide sufficient information for live user detection.

### 4.7 Overall Performance of Text-independent Liveness Detection

To secure conversational voice assistants and continuous voice authentication, we build a text-independent liveness detection solution based on phoneme templates. Figure 18 present the overall accuracy and EER of the text-independent liveness system. We could observe that, similar to the performance of the text-dependent liveness detection, the combined features result in the best performance with 97.65% accuracy and 3.52% EER. Whereas the energy-based features yield better performance than the frequency-based features, which, nevertheless, both achieve accuracy around 95%. Comparing with the text-dependent liveness detection, whose best accuracy is around 99%, our system realize text-independent liveness detection at a small cost of around 2% accuracy loss. Such slight degradation is normal as it could protect the whole conversation from replay attacks.

### 4.8 Overall Performance of Medium-range Liveness Detection on Smart Speakers

With the microphone array on smart speakers and the beamforming technique, we realize medium-range liveness detection to support various use cases in IoT environments. For example, in smartphone use cases, the users may hold their device within 30cm to themselves. In a smart vehicle, a driver could activate the auto-drive function by talking with the in-car Audio System around 50cm away. Whereas in smart home environments, a user may sit on the couch while interacting with a smart speaker on the end table locating at a distance of 100cm. To start the experiment, we connect the UMA-8 to a laptop in an office environment. We ask one participant to stand in front of the microphone array with distances of 0 cm, 20 cm, 40 cm, 60 cm, 80 cm, and 100 cm. The participant is asked to speak 3 sentences at each distance, and to repeat each sentence for 10 times. During the experiments, we use a smartphone (Note5) to emit the 20 kHz probe signal, and record the reflected probe signal, as well as the voices with the microphone array at a sampling frequency of 48 kHz. Figure 19 shows the liveness detection accuracies of our solution on the UMA-8 smart speaker with and without beamforming. We could observe that when the user is 20 cm away, the accuracy resulted from single microphone recordings drops to as low as 89.55%. Moreover, with increasing distance, the accuracy keeps dropping and jumps to 76.49% at 60 cm, 71.26% at 80 cm, and 41.87% at 100 cm. In comparison, the beamformed audio recording results in stable high accuracy above 98% for all test distances.

## 5 DISCUSSION

**Unconventional Loudspeaker.** In our work, we have tested conventional loudspeakers including the standalone speakers, the built-in speakers of mobile devices, and the earbuds. Nevertheless, there exists unconventional loudspeakers that do not relies on the diaphragm movement for sound production. For example, a piezoelectric speaker relies on a ceramic disc that interacts when it feels a certain voltage difference. A higher signal amplitude VPP (Voltage peak to peak) results in a larger piezo deformation and leads to a larger volume. Such a mechanism is fundamentally different from human speech production system. Another example of unconventional loudspeaker is the Electrostatic Loudspeaker (ESL), which still relies on the diaphragm movements for sound production. It is however, driven by two metal grids or stators instead of voice coil. As our liveness detection system relies on the movements of articulators for live user detection. Playing back with such loudspeakers can still be detected as a replay attack.

**Individual Diversity.** In our evaluation, we have tested our system when an attacker mimics the articulatory gesture of a genuine user by observing how the user pronouncing the passphrase. We now show how the performance looks like when an attacker has no prior-knowledge on how the legitimate user speaks. That is, the attacker use his own way of pronouncing the passphrase. This case is equipotent to compare the Doppler shifts of the articulatory gesture between two people who speak the same passphrase with their own habitual ways. Figure 20 shows the accuracy comparison. We observe that we could be able to achieve much higher accuracy at close to 100%. The result demonstrates that it is relative easier to capture the individual diversity than that of a mimicry attack.

**Limitations.** Our system is evaluated with a limited number of young and educated subjects. It will be useful to evaluate the system with a larger number of participants with a more diverse background to better understand the performance. Moreover, the system is evaluated only for several months. A long-term study could be conducted to consider the case that the individual characteristics is likely to change over lifetime. Nevertheless, we believe updating user profile periodically could potentially mitigate such a limitation.

## 6 RELATED WORK

Although the number of mobile applications that use voice biometric for authentication is rapidly growing, recent studies show that voice biometrics is vulnerable to spoofing attacks [13]. Acoustic feature based methods for attack detection have wide applicability, but they all have very limited effectiveness [22]. Current commercial voice authentication system like VoiceVault and Nuance, mostly rely on the challenge-response based methods to detect replay attacks. Such methods however require explicit user cooperation in addition to standard voice authentication process, which could be cumbersome. Many smartphone based solutions require the user to hold or move the phone in some predefined manners. For example, Zhang *et al.* measure the phonemes' time-difference-of-arrival (TDoA) dynamics to the two microphones of the phone when a live user speaks for liveness detection [28]. Though effective, it necessitate the users to hold the smartphone in front of their mouths. Moreover, many liveness detection solutions are only effective when the user and the device are in close proximity. For instance, Wang *et al.* [24] require the users to hold the phone close to their mouth and detect the human speakers' featured breathing sounds for liveness detection. Whereas Shang *et al.* [20] and Wang *et al.* [23] ask the users to hold the smartphone against their throats and chests to examine the throat vibrations and heart beats respectively for liveness detection. Furthermore, some researchers resort to extra devices for liveness detection. One example is the WiVo system that quantifies the Wi-Fi signals' CSI (Channel State Information) changes caused by mouth motions [17]. Similarly, REVOLT uses Wi-Fi to measure the breathing rate [19], whereas VocalPrint employs mmWave to sense vocal vibrations [16]. Besides, VAuth collects body vibrations with wearable sensors for liveness detection [11]. 2MA asks the users to prove their presence with similar recordings collected from multiple devices carried or close to the user [10]. Yan *et al.* require two spaced microphones to measure the field prints. The effectiveness of this method could be largely affected by the size of the device (distance between the microphones) [26].

In contrast, our system is transparent to users and covers more user cases as it works when holding the phones either to the user's ears or in front of their mouths. Moreover, our system is less susceptible to environmental noises as it senses articulatory gestures by actively emitting high frequency sound waves (which could be easily separated from noises) as oppose to passively listen to the voices that mixed with background noises in VoiceLive.

## 7 CONCLUSIONS

In this paper, we developed a voice liveness detection system requiring only a speaker and a microphone that are commonly available on smart devices. Our system, VoiceGesture, is practical as no cumbersome operations are required besides the conventional voice authentication process. Once it is integrated with voice authentication system, the liveness detection is transparent to the users. VoiceGesture performs liveness detection by extracting Doppler shift features caused by the articulatory gesture when a user speaks. Extensive experimental evaluation demonstrates the effectiveness of our system under various conditions, such as with different device types, placements and sampling rates. Moreover, VoiceGesture supports medium range liveness detection in various smart speaker use cases in smart homes and smart vehicles. Overall, VoiceGesture can achieve over 99% accuracy, with the EER at around 1% for text-dependent liveness detection, whereas around 98% accuracy and 3% EER for text-independent liveness detection.